# A consistent second-order hydrodynamic model in the time domain for floating structures with large horizontal motions

Yanlin Shao[1] | Zhiping Zheng[2] | Hui Liang[3] | Jikang Chen[2]

[1]Department of Mechanical Engineering, Technical University of Denmark, 2800, Lyngby, Denmark

[2]College of Shipbuilding Engineering, Harbin Engineering University, 150001, Harbin, China

[3]Technology Centre for Offshore and Marine, 118411, Singapore

**Correspondence**
Room 128, Building 403, Anker Engelunds Vej 1, Technical University of Denmark, 2800 Lyngby, Denmark
Email: yshao@mek.dtu.dk

**Funding information**
Independent Research Fund Denmark, Grant/Award Number: 013600329B

**ABSTRACT**

Floating offshore structures often exhibit low-frequency oscillatory motions in the horizontal plane, with amplitudes in the same order as their characteristic dimensions and larger than the corresponding wave-frequency responses, making the traditional formulations in an inertial coordinate system inconsistent and less applicable. To address this issue, we extend and explore an alternative formulation completely based on a non-inertial body-fixed coordinate system. Unlike the traditional seakeeping models, this formulation consistently allows for large-amplitude horizontal motions. A numerical model based on a higher-order boundary element is applied to solve the resulting boundary-value problems in the time domain. A recently developed new set of explicit time-integration methods, which do not necessitate the use of upwind schemes for spatial derivatives, are adopted to deal with the convective-type free-surface conditions. To suppress the weak saw-tooth instabilities on the free surface in time marching, we also present novel low-pass filters based on optimized weighted-least-squares, which are in principle applicable for both structured and unstructured meshes. The presented schemes for the convection equation and the low-pass filter are also relevant for other engineering fields dealing with similar mathematical problems. For ship seakeeping and added resistance analyses, we show that the present computational model does not need to use soft-springs for surge and sway, in contrast to the traditional models. For a floating monopile, the importance of consistently taking into account the effects of large horizontal motions is demonstrated considering the bi-chromatic incident waves. The present model is considered as a complete 2nd order wave-load model, as all the 2nd order wave loads, including the sum-frequency and difference-frequency components, are solved simultaneously.

## 1. INTRODUCTION

Floating structures are often the preferred solutions as supporting infrastructures in the industries of offshore oil and gas, marine renewable, and marine aquaculture engineering, at sites with deeper water depths, where bottom-fixed structures become excessively expensive. They are more cost-attractive as the weights of the superstructures are partly supported by the buoyancy of surrounding water, instead of entirely by the steel or concrete materials. The majority of offshore floating structures are moored by mooring systems that are anchored to the seafloor.

A mooring system provides restoring loads to a floater when it is offset from its static equilibrium position due to waves and currents, etc. in the oceans. To avoid possible resonant motions excited by wave-frequency (WF) loads, a typical mooring design will lead to large natural periods in the horizontal plane for offshore structures, in the order of 60 seconds or higher (Faltinsen, 1993). Since the waves normally have very little energy at such high wave periods, depending on the wave spectrum of the sea states of interest, the horizontal motions induced by the WF loads can be relatively small. However, due to the presence of non-linearities in the incident waves and their nonlinear interaction with the floaters, low-frequency (LF) wave loads can excite resonant horizontal motions of the floaters. Since the hydrodynamic system has very small wave-radiation damping at such high wave periods, the amplitudes of those resonant LF motions are often very large for typical offshore structures, and thus potentially threaten the integrity of the mooring systems.

To model the nonlinear wave loads and the hydrodynamic responses of the offshore structures, advanced fully-nonlinear models exist, such as the time-domain fully-nonlinear potential flow models, the Navier-Stokes solvers with proper turbulence modeling, or other equivalent models such as Smoothed-Particle Hydrodynamics (e.g. Leonardi et al., 2016) and Lattice Boltzmann Method (e.g. De Rosis et al., 2020). Among others, the fully-nonlinear potential flow models include those based on Boundary Element Method (Grilli, 1997; Ferrant et al., 2003; Bai and Eatock Taylor, 2006 2007 2009; Zhou et al., 2016; Zhang and Teng, 2021), Finite Element Method (Tou, 1991; Wu and Eatock Taylor, 2003; Wang and Wu, 2006 2007; Yan and Ma, 2007; Sun et al., 2015; Huang and Wang, 2020), Finite Difference Method (Bingham and Zhang, 2007; Engsig-Karup et al., 2009; Ducroze et al., 2014; Hicks et al., 2021), and Harmonic Polynomial Cell method (Shao and Faltinsen, 2012 2014a; Hanssen et al., 2018; Liang et al. 2020; Tong et al., 2021). Navier-Stokes solvers have also become widely used in various hydrodynamic analyses for offshore structures. Some of them are preferred in the academic and engineering communities. Examples are the open-source tool OpenFOAM (Jacobsen et al. 2012; Higuera et al., 2013; Hu et al., 2016) and the commercial software STAR-CCM+ (Oggiano et al., 2017).

The above-mentioned fully-nonlinear methods are still considered too time-consuming to be heavily applied in the design of offshore structures, even though significant progress has been made in the past decades. The leading order nonlinear effects governing the slow-drift LF response of floating structures depend quadratically on the wave amplitudes. For irregular waves, a complete account of all the 2nd order wave effects requires the solutions of the 1st, 2nd, and 3rd order free-surface problems, as discussed in Sclavounos (1992). With a typical wave spectrum, however, the LF quadratic effects are dominated by the 1st and 2nd order free-surface problems, even though the 3rd order solution may also be needed in the complementary high-frequency regime. It is fair to say that the state-of-the-art tools applied in the related industries to simulate the LF slow-drift motions of offshore structures are based on the 1st and 2nd order potential-flow weakly-nonlinear theories. Due to the limitation of potential flow assumptions, viscous effects which are considered secondary for large-volume offshore structures, are normally accounted for empirically when necessary (Faltinsen, 1993).

Within the context of 2nd order weakly-nonlinear theory, the associated wave radiation and diffraction problems have been extensively studied by analytical and numerical means, with some pioneering work started in the 1970s (Faltinsen and Løken, 1979; Lighthill, 1979; Molin, 1979). Semi-analytical solutions have been developed for structures with simple geometries, e.g. bottom-mounted and truncated vertical circular cylinders (Eatock Taylor and Hung, 1987; Eatock Taylor and Chau, 1992; Huang and Eatock Taylor, 1996). From a practical application point of view, it is of great interest to develop accurate and efficient numerical methods to analyze offshore structures of arbitrary geometries. In this regard, both frequency-domain and time-domain models have been developed in the past decades. Frequency-domain models can be found, for instance, in Molin (1979), Lee and Newman (1994), Chen et al. (1995), Newman and Lee (2002), Huang et al. (2021), and Cong et al. (2021). Some of those efforts have laid a good basis in the implementation of start-of-the-art frequency-domain radiation/diffraction solvers, e.g. WAMIT and HydroStar.

Another branch of the development with success is in the time domain, where it is more convenient to include other transient or nonlinear effects, for instance, nonlinear restoring forces from a mooring system or viscous-drag loads on slender components of the floating system. Similar to as it has been done in the frequency domain, it has become a common practice to solve the hydrodynamic problem in the Earth-fixed reference frame. Some excellent examples are found in, among others, Isaacson and Ng (1993), Skourup et al. (2000), Teng et al. (2002), Wang and Wu (2007), Bai and Teng (2013), and Huang and Wang (2020).

There are at least two challenges in applying the traditional formulation based on the inertial coordinate system (*IneCS*) in studying any floating structures in ocean waves: (i) A traditional formulation in *IneCS*, assuming the 2nd order motions much smaller than the corresponding 1st order components, becomes inconsistent when dealing with large horizontal motions. In practice, the LF horizontal motions of a floating system maybe even larger than not only the 1st order motions but also the characteristic dimensions of a floater. (ii) The body-boundary conditions contain higher derivatives of the velocity potential, which are very difficult to calculate accurately, or not integrable close to the sharp edges of the offshore structures.

To address the first challenge, Teng et al. (2016) were perhaps the first in the literature to extend the *IneCS* model by Taylor-expanding the boundary conditions around an instantaneous mean position of the structure, with a cost of approximating the free-surface variables at the instantaneous position, e.g. velocity potential and wave elevation, by additional Taylor expansions involving the evaluation of more higher-derivative terms. The second challenge was also partly resolved numerically by Teng et al. (2016) to use Stokes theorem to convert the surface-integral of 2nd derivatives of velocity potential into 1st derivatives, with an extra cost of increased order of the derivatives of Green's function. It is however not clear how the 3rd derivatives of the zeroth-order potential, which according to Shao and Faltinsen (2013) should be present in the 2nd order boundary conditions, can be handled by a similar strategy. Furthermore, the higher derivatives still need to be evaluated when calculating the wave loads if a pressure-integration approach is applied.

An alternative mathematical formulation in the body-fixed coordinate system (*BFCS*) has been proposed by Shao and Faltinsen (2010, 2013) as a general framework to deal with linear and weakly-nonlinear wave-structure interaction problems. It avoids calculation of higher derivatives on the wetted surface of the structures, thus resolves the above-mentioned second challenge theoretically. The advantage of using this formulation has been demonstrated in the analyses of linear and 2nd order sum-frequency wave loads (Shao and Faltinsen, 2010 2013 2014b; Zheng et al., 2020) on ships and offshore structures. However, its great advantage, as will be shown later in the paper, has not been explored to address the above-mentioned first challenge in modeling floating offshore structures with large horizontal motions.

Our contributions in this paper are at least threefold. Firstly, the *BFCS* formulation will be extended to address the challenges in the hydrodynamic analysis of floating structures with large horizontal motions. Since the formulation does not assume the horizontal motions as small variables, it does not have the same inconsistency as in a traditional formulation in *IneCS*. As the horizontal LF velocities of the floating structure are considered as quasi-steady speeds, the free-surface conditions have similarities to the convective equations. The numerical model developed in this paper can consistently handle scenarios involving slowly-varying horizontal displacement and yaw angles, which is beyond the capability of the model in Zheng et al. (2020). In fact, the present model is applicable for arbitrarily large horizontal displacements, and yaw angles, e.g. up to 360 degrees. Secondly, to stabilize the time-domain solution of those equations, efficient explicit schemes, approximated from the corresponding implicit schemes, will be presented for general convection equations. Compared with Zheng et al. (2020), we start with a weighted-average formulation in the spatial discretization of the convection equations and derive more general equations for the approximated schemes. In contrast to some other explicit schemes, which often need upwind finite-difference to approximate the spatial derivatives in the convective terms, those schemes are conditionally stable even if central-difference schemes are applied. The matrix-based stability analyses will also be presented in this paper to rigorously support the correctness and effectiveness of the developed schemes. Thirdly, considering that the most commonly used low-pass filters have been developed for structured meshes, we have developed novel low-pass filters based on optimized weighted-least-squares (WLS), which are applicable for unstructured meshes as well. Comparison with the Savizky-Golay filters in one dimension and the conventional two-dimensional (2D) least-squares filters show that the new filters are more effective in eliminating short waves and less diffusive for longer waves.

Those relatively new theoretical and numerical developments, together with a higher-order boundary element method (BEM) based on cubic shape functions as a solver for the Laplace equation, present a unique time-domain computational model, which is complete and consistent up to 2nd order, for the analysis of floating offshore structures with large horizontal motions. To demonstrate the capacity of the new model, two practical applications will be presented: The first concerns the linear WF ship seakeeping and added resistance in the time domain, for which the current best-practice models need to use artificial soft-springs to avoid the failure of the solutions. Thanks to the adopted special mathematical formulation, as will be demonstrated later, our model does not need soft-springs for the horizontal surge and sway motions. In the second application, we study the rigid-body motions of a floating monopile in bi-chromatic waves, where surge resonance occurs due to the 2nd order LF excitation loads. The importance of consistently accounting for the large-amplitude motions will be demonstrated.

The remainder of this paper is organized as follows. In Section 2, the formulation of the 1st and 2nd order boundary-value problems in the body-fixed coordinate system is presented. Section 3 briefly introduces the higher-order BEM, which will be applied in solving the Laplace equation. In Section 4, a new set of conditionally stable explicit time-integration methods, specially designed for convection equations, are presented, followed by the introduction of body-motion equations in Section 5. Section 6 provides the derivation of the new low-pass filters based on the optimized weighted-least-square procedure, which are applicable for both structured and unstructured meshes. Section 7 presents two practical applications, one for seakeeping and added resistance of the KVLCC2 ship where extensive validation and verification materials are available, and the other for the motion of a floating monopile under the action of bi-chromatic waves. Section 8 summarizes the present work.

## 2. MATHEMATICAL FORMULATION

The main assumption in our mathematical formulation is that the potential-flow theory can be used, which means we consider inviscid and incompressible flows without any rotational motions. Therefore, the governing equation in three dimensions is:

$$\nabla^2 \phi = 0. \tag{1}$$

Here $\phi$ is the velocity potential, whose derivatives are the fluid velocities. Earlier examples of potential-flow applications in marine hydrodynamics can be found in, for instance, von Karman (1929) and Wagner (1932). More general descriptions can also be found in textbooks, for instance, Faltinsen (1993).

To facilitate the definition, we define three different coordinate systems, as shown in Figure 1, including an Earth-fixed coordinate system $O_e X_e Y_e Z_e$, $OXYZ$ moving with the steady or LF motions in the horizontal plane but not the WF motions, and a body-fixed coordinate system (*BFCS*) $oxyz$, in which the mathematical formulation of the boundary-value problem (BVP) will be defined. $OXYZ$ is an inertial reference frame if it moves with a steady speed, and it becomes non-inertial if it moves with a slowly-varying speed, e.g. the slow-drift velocity of a floating structure. The main features of the three different coordinate systems are summarized in Table 1. Unless otherwise mentioned, all variables will be described in the non-inertial coordinate system $oxyz$ hereafter.

**TABLE 1.** Main features of the three coordinate systems.

|  | $O_e X_e Y_e Z_e$ | $OXYZ$ | $oxyz$ |
| --- | --- | --- | --- |
| Inertial | Yes | Maybe | No |
| Moving vel. | Earth-fixed | LF vel. | LF + WF vel. |

The boundary conditions on the free surface without considering the surface-tension effects in $BFCS$ ($oxyz$ in Figure 1) have been derived in, for instance, the textbook by Faltinsen and Timokha (2009) as

$$\eta_t = \phi_z - \overline{\nabla}\phi \cdot \overline{\nabla}\eta - (\boldsymbol{W} + \boldsymbol{v} + \boldsymbol{\omega} \times \boldsymbol{r}) \cdot (-\eta_x, -\eta_y, 1) \quad (2)$$

$$\phi_t = -\frac{1}{2}\nabla\phi \cdot \nabla\phi + (\boldsymbol{W} + \boldsymbol{v} + \boldsymbol{\omega} \times \boldsymbol{r}) \cdot \nabla\phi - U_g. \quad (3)$$

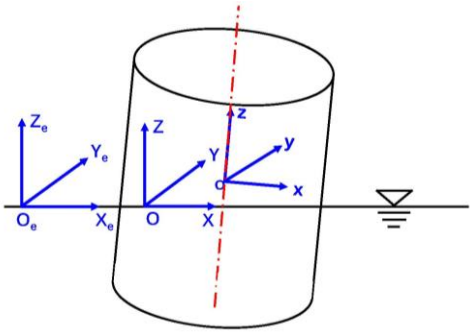

**FIGURE 1**. Definition of the three coordinate systems: $O_e X_e Y_e Z_e$, $OXYZ$ and $oxyz$.

Equation (1) is the kinematic free surface, which requires that a fluid particle on the free surface will stay on the free surface all the time. Equation (2) is the dynamic free surface, requiring the pressure to be the same as atmospheric pressure, which is taken as zero since its value does not have any influence on the hydrodynamic results. Both conditions are satisfied on the instantaneous free surface $z = \eta(x, y, t)$, where $\eta$ is the instantaneous free surface elevation. $\boldsymbol{W}$ is the steady or LF velocity of a fixed point in $BFCS$ due to the translational and rotational motions of the coordinate system. $\boldsymbol{v} = (\dot{\xi}_1, \dot{\xi}_2, \dot{\xi}_3)$ is the unsteady translational velocity vector of the structure. $\boldsymbol{\omega} = (\dot{\alpha}_4, \dot{\alpha}_5, \dot{\alpha}_6)$ represents the unsteady angular velocity. $\boldsymbol{r} = (x, y, \eta)$ denotes the position on the instantaneous free surface. $\nabla$ is the three-dimensional (3D) gradient operator in $oxyz$, and $\overline{\nabla}$ is a 2D gradient operator on $xy$- plane. $U_g = -\boldsymbol{g} \cdot \boldsymbol{r}$ represents gravitational potential evaluated in $BFCS$, with $\boldsymbol{g}$ as the gravity acceleration vector. A subscript in the equations means a partial differentiation.

Solving the free-boundary problem governed by the Laplace equation, the fully-nonlinear free-surface conditions in Equations (2) and (3), and the body-boundary condition, is very time-consuming. The solution will also fail when the wave breaks, thus the fully nonlinear potential-flow model is not yet ready as a suitable tool for engineering design. Therefore, linear and weakly-nonlinear models based on perturbation schemes, which are much more computationally efficient, have become popular in the computer-aided design of offshore structures, in particular in the analysis of WF and LF motions, as well as mooring loads. Following the procedure of the perturbation scheme, one can Taylor-expand the free-surface conditions around $xy$- plane, introduce the Stokes expansion for velocity potential $\phi$, wave elevation $\eta$, and WF body motions, and consistently collect terms at the same order. Different from a description in $IneCS$, the observed wave elevation $\eta(x, y, t)$ in $BFCS$ has a contribution from the rigid-body motion in the vertical plane, which is known when the body motions have been solved or prescribed. Therefore, the following decomposition has been introduced to facilitate the numerical analysis (Shao and Faltinsen, 2010 2013).

$$\eta = \tilde{\eta} + \bar{\eta} + O(\epsilon^3). \quad (4)$$

Here $\epsilon$ is a parameter, e.g. the wave steepness, to denote the smallness of the variables. $\tilde{\eta}$ is the unknown part of the total wave elevation $\eta(x, y, t)$ due to diffraction and radiation effects, which will be solved as part of the solution. The final expressions for the 1st and 2nd order free-surface conditions can be written as

$$\left[\frac{\partial}{\partial t} - (\boldsymbol{W}^{(0)} - \overline{\nabla}\phi^{(0)}) \cdot \overline{\nabla}\right]\phi^{(k)} = f_1^{(k)}, \quad (5)$$

$$\left[\frac{\partial}{\partial t} - (\boldsymbol{W}^{(0)} - \overline{\nabla}\phi^{(0)}) \cdot \overline{\nabla}\right]\tilde{\eta}^{(k)} = f_2^{(k)}. \quad (6)$$

Here we have used a superscript $(k), k = 1, 2$ to denote the 1st and 2nd quantities. Both Equations (5) and (6) are satisfied on $xy$-plane, defined on the plane with $z = 0$. The forcing terms $f_1^{(k)}$ and $f_2^{(k)}$ in Equations (5) and (6) are

$$f_1^{(1)} = -g\tilde{\eta}^{(1)} + (\boldsymbol{\chi}^{(1)} + \boldsymbol{W}^{(1)}) \cdot \nabla\phi^{(0)}, \quad (7)$$

$$\begin{aligned}f_1^{(2)} &= -g\tilde{\eta}^{(2)} - \tfrac{1}{2}\nabla\phi^{(1)} \cdot \nabla\phi^{(1)} \\ &\quad -\eta^{(1)}\left(\phi_{tz}^{(1)} + \nabla\phi_z^{(1)} \cdot \nabla\phi^{(0)} + \nabla\phi_z^{(0)} \cdot \nabla\phi^{(1)} - \boldsymbol{W}^{(0)} \cdot \nabla\phi_z^{(1)}\right) \\ &\quad +(\boldsymbol{\chi}^{(2)} + \boldsymbol{W}^{(2)}) \cdot \nabla\phi^{(0)} + (\boldsymbol{\chi}^{(1)} + \boldsymbol{W}^{(1)}) \cdot \left(\nabla\phi^{(1)} + \eta^{(1)}\nabla\phi_z^{(0)}\right),\end{aligned} \quad (8)$$

$$f_2^{(1)} = \phi_z^{(1)} + \eta^{(1)}\phi_{zz}^{(0)} - \overline{\nabla}\phi^{(0)} \cdot \overline{\nabla}\bar{\eta}^{(1)}, \quad (9)$$

$$\begin{aligned}f_2^{(2)} &= \phi_z^{(2)} + \eta^{(2)}\phi_{zz}^{(0)} + \eta^{(1)}\phi_{zz}^{(1)} + (\boldsymbol{W}^{(0)} - \overline{\nabla}\phi^{(0)}) \cdot \overline{\nabla}\bar{\eta}^{(2)} \\ &\quad -\overline{\nabla}\phi^{(1)} \cdot \overline{\nabla}\eta^{(1)} + \boldsymbol{W}^{(1)} \cdot \overline{\nabla}\bar{\eta}^{(1)} - \boldsymbol{W}_3^{(2)} + \boldsymbol{\chi}^{(1)} \cdot \overline{\nabla}\tilde{\eta}^{(1)}.\end{aligned} \quad (10)$$

Here $\boldsymbol{W}^{(0)}$ denotes a steady or LF velocity of the structure on the $xy$-plane. $\boldsymbol{W}^{(k)}, k = 1, 2$, is the $k$th order component of the steady or LF speed of the floating structure, observed in $BFCS$, i.e. $oxyz$ reference frame. If the steady or LF velocity in the horizontal plane is described in $OXYZ$ system as $\boldsymbol{U} = (U - \Omega y)\boldsymbol{I} + (V + \Omega x)\boldsymbol{J} + 0\boldsymbol{K}$, which is often the case in offshore applications, $\boldsymbol{W}^{(k)}$ can be obtained by

$$\boldsymbol{W}^{(k)} = \mathcal{R}_{i \to b}^{(k)}\boldsymbol{U}, \quad k = 1, 2, \quad (11)$$

where $U, V, \Omega$ represent steady or LF velocities in the surge, sway, and yaw, respectively. $\boldsymbol{I}$, $\boldsymbol{J}$, and $\boldsymbol{K}$ are the unit vectors along $X$-, $Y$- and $Z$-axis, respectively. Note that we have used uppercase letters to denote the variables in the coordinate system $OXYZ$, as lowercase letters are reserved for $oxyz$ system. $\mathcal{R}_{i \to b}^{(k)}$ is the $k$th order component of the transformation matrix from $OXYZ$ to $oxyz$. The subscripts `$b$' represents the body-fixed reference frame $oxyz$, and `$i$' is used to denote $OXYZ$, even though it may not always be an inertial reference frame. Following the rotation in the sequence of $Z$-, $X$-, $Y$-axes (e.g., see Ogilvie, 1983; Shao, 2010), one can obtain the expressions of $\mathcal{R}_{i \to b}^{(k)}$ as a function of angular motions, which are included in the Appendix for the completeness of the paper. Details of derivation can be found in, among others, Ogilvie (1983) and Shao (2010). $\chi^{(k)}$, $k = 1, 2$, is the $k$th order rigid-body velocity of a point on $xy$-plane, which is defined as

$$\chi^{(k)} = v^{(k)} + \dot{\mathcal{R}}_{b \to i}^{(k)} r. \tag{12}$$

Here we have used an over-dot to denote time differentiation.

The free-surface conditions in (5) and (6) involve the zeroth-order potential $\phi^{(0)}$ and its derivatives, which are obtained by solving a double-body-flow problem (e.g., see Ogilvie and Tuck, 1969), satisfying Laplace equation in the fluid domain, $\partial \phi^{(0)}/\partial z = 0$ on $z = 0$, and $\partial \phi^{(0)}/\partial n = \boldsymbol{n} \cdot \boldsymbol{U}$ on the mean wetted surface of the structure. As seen from Equations (5)-(10), most of the terms in the free-surface conditions of the *BFCS* model are similar to that in the traditional *IneCS* formulation, while the underlined terms only appear in the *BFCS* model. Those terms do not increase the difficulties in the numerical evaluation of the right-hand sides of the free-surface conditions, since the highest derivatives, $\phi_{zz}^{(0)}$ and $\phi_{zz}^{(1)}$ in Equations (9) and (10) also need to be calculated in an *IneCS* formulation.

The great advantage of the *BFCS* formulation is the simplicity of the body-boundary conditions. The 1st and 2nd order body-boundary conditions in *BFCS* are

$$\frac{\partial \phi^{(k)}}{\partial n} = \boldsymbol{n} \cdot \left( v^{(k)} + \omega^{(k)} \times r + \mathcal{R}_{i \to b}^{(k)} U \right), \quad k = 1, 2, \text{ on } S_B, \tag{13}$$

where $S_B$ denotes the mean wetted body surface. $\boldsymbol{n}$ is the normal vector on the body surface, described in *BFCS*. The last term on the right-hand side represents the angle-of-attack effects due to the forward speed (Faltinsen, 1993).

The *BFCS* formulation of the body-boundary conditions in Equation (13) does not involve evaluation of the higher derivatives on the body surface $S_B$, since a Taylor expansion was not needed in the derivation. Therefore, the *BFCS* model mathematically avoids the inconsistency and numerical difficulties associated with the higher derivatives on the body surface, for instance, the so-called $m$-terms $\left[ \boldsymbol{x}^{(k)} \cdot \nabla(\nabla \phi^{(0)}) \right] \cdot \boldsymbol{n}$ described in the *IneCS* models (e.g., see Zhao and Faltinsen, 1989a 1989b; Wu, 1991; Kring, 1994; Chen and Malenica, 1998; Shao, 2010). Here $\boldsymbol{x}^{(k)}$ is the $k$th order unsteady motion on the body surface. For moored offshore structures with large horizontal motions, those terms potentially lead to nonphysical or divergent numerical results. The vertical motions (heave, roll, and pitch) are not problematic in this regard, as hydrostatic restoring loads are normally non-negligible for surface-piercing structures, thus the corresponding vertical motions are bounded. In the time-domain seakeeping analysis of ships, for which there might be no restoring forces/moments for surge, sway, and yaw motions, artificial soft-springs are commonly applied for those motions (see, e.g., Kring, 1994; Seo et al., 2014; Kim et al., 2017), even in the case that only the steady-state WF responses in regular waves are sought for. In Section 7, we will show that the *BFCS* model does not have to apply soft-springs for surge and sway to obtain stable steady-state WF responses of ships in regular waves. The reason is that the *BFCS* model does not involve the surge and sway displacements in the formulation, but rather only the corresponding velocities. Those horizontal displacements are only needed when updating the incident wave fields around the structure surface and the free surface surrounding the structure.

In this section, we have referred a few times to the traditional *IneCS* formulation. However, due to limited space, the boundary conditions on the free surface and body surface in an *IneCS* formulation will not be repeated here, as it is not our focus in this paper. We refer interested readers to, for instance, Bai and Teng (2013) and Teng et al. (2016), for a comprehensive description of this formulation up to 2nd order.

The initial boundary-value problem defined in this section will be solved in the time domain, which will involve several keys steps, including accurate solution of the 3D Laplace equation, stable time-integration of the free-surface conditions, low-pass filtering of the solution on the free surface and time-integration of the rigid-body motion equations. As an illustration, a simple flow-chart of the overall solution procedure is given in Figure 2. More details of the numerical schemes for each step will be explained in the following sections.

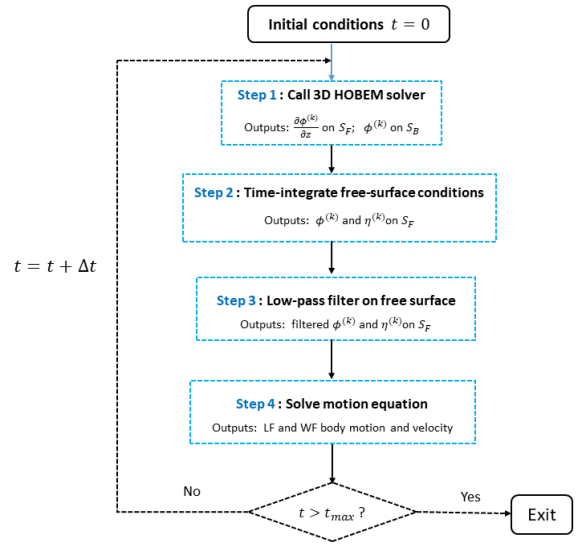

**FIGURE 2**. Flow-chart of the solution steps in the present numerical model.

## 3. NUMERICAL SOLUTIONS OF THE 3D LAPLACE EQUATION

The numerical solutions of the BVPs defined for $\phi^{(k)}$, $k = 1, 2$, in Section 2 will be obtained by a 3D Laplace equation solver. In this study, a higher-order BEM is used, which implies that Green's 3rd identity will be applied to convert the 3D Laplace equation into a boundary integral equation, which only needs discretization on the boundaries of the fluid but not in the whole fluid domain.

In the present study, the fluid boundaries are discretized by piecewise 12-node elements (or patches). An example of a 12-node cubic patch is shown in Figure 3. The left sub-figure shows its representation in the physical space $xyz$, and its representation on the parametric $\xi\eta$-plane is shown in the right sub-figure. Using the 12-node elements, we can interpolate the geometries represented by $(x,y,z)$, the velocity potential $\phi$, and its normal derivatives $\frac{\partial \phi}{\partial n}$, etc. by the values at the 12 nodes as

$$x = x_j N_j(\xi,\eta); \quad y = y_j N_j(\xi,\eta); \quad z = z_j N_j(\xi,\eta), \quad (14)$$

$$\phi = \phi_j N_j(\xi,\eta); \quad \frac{\partial \phi}{\partial n} = \left(\frac{\partial \phi}{\partial n}\right)_j N_j(\xi,\eta), \quad (15)$$

where $N_j(\xi,\eta), j = 1,\ldots,12$, is the $j$th shape function (e.g., see Shao, 2010 for detailed expressions). The Einstein summation notation has been used in Equations (14) and (15), and thus a duplicate index $j$ means summation. The spatial derivatives of the physical variables, for instance, the velocity potential and wave elevation, can be obtained with the help of the higher-order shape functions. Specifically, to get the spatial derivative of a physical variable at a point shared by several elements, we calculate the spatial derivatives on all elements sharing this point, and an average of them is then taken. This approach is similar to the central-difference type of differentiation. More details of the applied higher-order BEMs based on shape functions can be found in, for instance, Shao (2010), Heo and Kashiwagi (2019), and Zheng et al. (2020). There are many different BEM solvers or boundary-element algorithms that could be used to solve the Laplace equation. It was decided to use the cubic BEM in our studies since the 2nd derivatives of the velocity potential are needed in the 2nd order free surface conditions. A lower-order BEM would need much more elements to achieve the same accuracy as a cubic order BEM.

At the waterline where the free surface and body surface intersect with each other, both the Dirichlet boundary condition from the free surface and the Neumann boundary condition from the body surface have been satisfied in the present BEM model. At each time step, the Dirichlet condition on the free surface is obtained through the time-integration of the free surface condition (see Section 4), and the Neumann condition on the body surface is obtained through the time-integration of rigid-body motion equations (see Section 5).

Since both the linear and 2nd order boundary conditions are satisfied on surfaces that do not vary in time, e.g. $xy$-plane for free surface condition and mean wetted body surface for body boundary condition, the matrix on the left-hand side of the resulting system does not change in time, thus it only needs to be inverted for once and its inverse can be used throughout the simulation. Therefore, the time-domain solvers based on the same or similar 2nd order theory in this paper are very time efficient compared with other fully nonlinear solvers. In this paper, all the simulations that will be presented in Section 7 were performed on a desktop PC with a 2.7 Hz Intel CPU.

The accuracy of the applied cubic BEM solver has been verified in some previous studies (see, e.g., 2010 2013 2014b), and thus will not be repeated in the present paper due to limited space and different focuses.

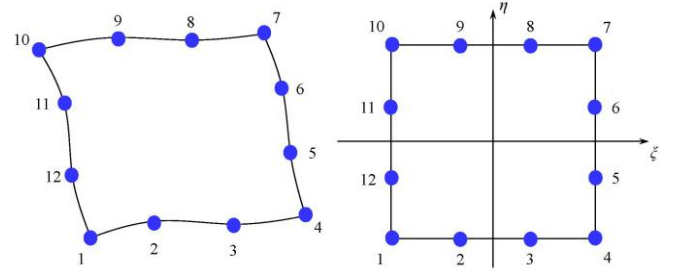

**FIGURE 3**. Representation of a 12-node cubic patch. Left: in physical space $xyz$; Right: on a parametric $\xi\eta$ plane.

## 4. TIME-INTEGRATION OF FREE SURFACE CONDITIONS

The free-surface conditions (5) and (6) will be integrated with respect to time to obtain the wave elevation and velocity potential at new time steps. Both equations can be regarded as convection equations with forcing (or source) terms on their right-hand sides, and they are also coupled with each other as seen from the forcing terms in Equations (7) - (10). In the literature, there are typically two popular ways to solve the convective equations: (i) Conditionally-stable explicit time-integration methods, e.g. explicit Euler or Runge-Kutta schemes, with upwind-difference (UD) schemes for spatial discretization, and (ii) Unconditionally-stable implicit schemes, e.g. the implicit Crank-Nicolson scheme. Other examples of popular time-integration schemes can be found in, for instance, Adeli et al. (1978) and Naess and Moan (2012). With the presence of oscillatory LF horizontal motions, the amplitude and/or direction of convection velocities at the free-surface mesh points will change as the solutions evolve in time. Thus, the use of explicit methods will necessitate searching for new upstream stencil points for each free-surface point at each time step, which is a non-trivial task in particular when unstructured meshes are adopted. Inspired by the applications in finite element methods based on unstructured meshes, Shao (2019) has used upwind differentiation along the streamlines in their time-domain BEM solutions. On the other hand, the implicit schemes can also be time-consuming, since a time-dependent sparse-matrix equation must be set up and solved at each time step. The entries of the sparse matrix will change due to the change of amplitude and/or direction of the convective velocities. In this case, the new set of explicit time-integration schemes that will be presented in the rest of this section seem to be good choices, as they allow for central difference for spatial derivatives and do not need to solve extra sparse-matrix equations. Those explicit schemes are approximated from implicit schemes and are conditionally stable.

Equations (5) and (6) can be formally written in form of a convection-type equation

$$\frac{\partial \psi}{\partial t} + \boldsymbol{c} \cdot \overline{\nabla} \psi = f(x,y,t), \tag{16}$$

where $\psi$ can be either the velocity potential or wave elevation. $\boldsymbol{c} = -(\boldsymbol{W}^{(0)} - \overline{\nabla} \phi^{(0)})$ is the convection velocity, with an amplitude of $c_m$. Let's discretize Equation (16) in a weighted-average manner

$$\frac{\psi^{n+1}-\psi^n}{\Delta t} + \theta c_m \mathbf{D} \psi^{n+1} + (1-\theta) c_m \mathbf{D} \psi^n = f^n, \ \theta \in [0,1]. \tag{17}$$

Here $\mathbf{D}$ is the directional derivative operator (matrix) to get $\frac{\boldsymbol{c} \cdot \overline{\nabla} \psi}{c_m}$ on all free-surface points. $\boldsymbol{c}/c_m$ is the unit vector along the direction of the convection velocity. Other bold symbols are vectors containing variables on the free surface. The superscripts $n$ and $n+1$ indicate the previous and current time steps, respectively. A weight factor $\theta \in [0,1]$ has been introduced. $\theta = 0$ leads to an explicit Euler method, while $\theta = 1$ means an implicit Euler method. The implicit Crank-Nicolson method can be obtained by letting $\theta = 0.5$.

For simplicity but without losing generality, we will ignore the forcing term $f^n$ in the derivation, as it does not influence the stability of the scheme. The effect of the forcing term $f^n$ can be easily included separately. Equation (17) can be rewritten as

$$\mathbf{A} \phi^{n+1} = \mathbf{B} \phi^n, \tag{18}$$

where $\mathbf{A} = \mathbf{I} + \Delta t \theta c_m \mathbf{D}$, and $\mathbf{B} = \mathbf{I} - \Delta t (1-\theta) c_m \mathbf{D}$. Here $\mathbf{I}$ is an identity matrix. The above sparse-matrix equation can be of course be solved numerically by a standard procedure. However, it is also possible to obtain an explicit approximation of the solution. Since the derivative operator $\mathbf{D}$ is inverse proportional to the element size, we rewrite matrix $\mathbf{A}$ as

$$\mathbf{A} = \mathbf{I} + \frac{\Delta t c_m \theta}{\Delta h} \mathbf{D} \Delta h \tag{19}$$

Here $\alpha = \Delta t c_m / \Delta h$ is the Courant–Friedrichs–Lewy (CFL) number which is normally taken to be less than 1.0 in explicit schemes. $\Delta h$ is a characteristic mesh size. It is introduced to non-dimensionalize the variables and does not influence the final equations that will be derived. With $\alpha$ as a small variable, we expand the inverse of $\mathbf{A}$ into a series expansion in $\alpha$ and require $\mathbf{A}\mathbf{A}^{-1} = \mathbf{I}$. Therefore, an explicit approximation of $\mathbf{A}^{-1}$ can be obtained as:

$$\mathbf{A}^{-1} = \mathbf{I} + \sum_{k=1}^{\infty} (-\theta \Delta t c_m \mathbf{D})^k. \tag{20}$$

Therefore, by substituting Equation (20) into Equation (18), and keeping only a finite number of $N_p$ terms in the summation, we can obtain the following explicit scheme as an approximation of the implicit scheme in Equation (18)

$$\boldsymbol{\phi}^{n+1} = \boldsymbol{\phi}^n + \Delta t \left[ \sum_{k=0}^{N_p-1} (-c_m \mathbf{D})^{k+1} (\Delta t \theta)^k \right] \boldsymbol{\phi}^n + O(\Delta t^2). \tag{21}$$

If only one term is kept in the truncated summation, i.e. $N_p = 1$, Equation (21) becomes the explicit Euler scheme. As $N_p \to \infty$, the scheme becomes the same as the original implicit scheme. In practice, since we normally operate with small Courant–Friedrichs–Lewy (CFL) numbers to achieve the desired accuracy, as we do in other standard explicit schemes, only a few terms are needed as the higher-order terms in CFL (or $\Delta t$) are less important.

To demonstrate the stability of the scheme described by Equation (21), we take a one-dimensional (1D) convection equation as an example and consider periodic boundary conditions at two ends of the domain. A matrix-based linear stability analysis has been carried out, and the results for $\theta = 0.5$ are shown in Figure 4 for four different CFL numbers, i.e. CFL = 0.5, 1.0, 1.5, and 2.0. The 2nd order central difference has been used to approximate the spatial derivative. Contrasting with many other explicit schemes where upwind difference has to be applied to approximate the spatial derivatives, the present scheme allows for the use of central-difference schemes. In Figure 4, $\Re(\lambda)$ and $\Im(\lambda)$ are the real and imaginary parts of the eigenvalues of the Jacobian matrix in the linear stability analysis, respectively. The time-integration scheme is considered stable if all the eigenvalues drop in the unit circle in the figure. 21 points have been used when discretizing the 1D computational domain. Different mesh points will lead to a different number of eigenvalues, but not change the conclusion of the present stability analysis.

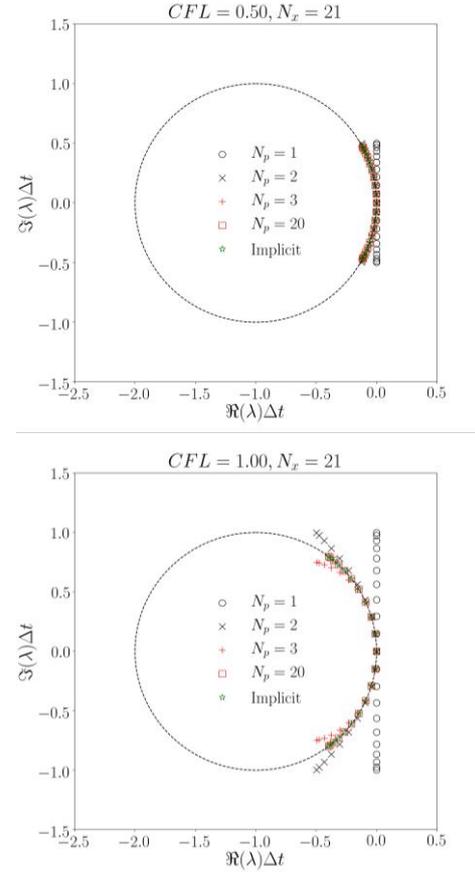

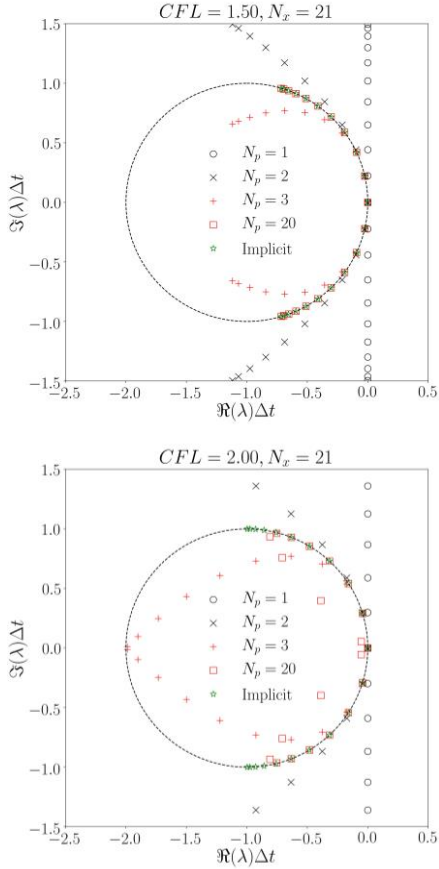

**FIGURE 4**. Stability region for the approximated implicit Crank-Nicolson scheme with $\theta = 0.5$ in Equation (21). Periodic boundary conditions are applied at inflow and outflow boundaries. The 2nd order central difference is used to approximate $\partial \phi / \partial x$. The results are shown for different $N_p$, which is the number of terms in the truncated series. Results for 4 different CFL numbers (0.5, 1.0, 1.5, 2.0) are presented.

If only one term is kept in the truncated summation in Equation (21), the scheme is unconditionally unstable as can be seen from the results of $N_p = 1$ in Figure 4. In this case, all eigenvalues are pure imaginary. When one extra term is included, i.e. $N_p = 2$, the scheme becomes stable up to CFL = 1.0, which however becomes unstable for larger CFL numbers. If two extra terms are included, i.e. $N_p = 3$, the scheme is still stable up to CFL = 2.0. In principle, the scheme can be stable for even larger CFL values, provided that enough number of terms are kept in the truncated summation in Equation (21), thus the scheme is closer to a fully implicit scheme. However, it is not beneficial to do so from a computational point of view. In practice, we only include very few terms in Equation (21), and $N_p = 5$ will be used for all the simulations that will be presented later in Section 7.

A numerical damping zone is applied at the outer layer of the truncated free-surface domain so that most of the energy of the outgoing waves has been damped out before they reach the end of the damping zone. The numerical damping zone is a standard approach in the time-domain simulation of wave-structure interaction to absorb outgoing waves, thus we will not repeat the details here. Interested readers are referred to, among others, Ferrant (1998), Ferrant et al. (2003), Bai and Eatock Taylor (2006), Shao and Faltinsen (2010 2012a), and Bai and Teng (2013).

## 5. TIME-DOMAIN SOLUTION OF THE MOTION EQUATIONS

The rigid-body motion equations may be formulated either in *IneCS* or *BFCS* reference frame. Here we use the following formulation in *BFCS*

$$m\left(\dot{\boldsymbol{v}}^{(k)} + \boldsymbol{\omega}^{(1)} \times \boldsymbol{v}^{(1)}\right) = \boldsymbol{F}_g^{(k)}, \tag{22a}$$

$$\boldsymbol{I}_B \dot{\boldsymbol{\omega}}^{(k)} + \boldsymbol{\omega}^{(1)} \times \left(\boldsymbol{I}_B \boldsymbol{\omega}^{(1)}\right) = \boldsymbol{M}_g^{(k)}, \tag{22b}$$

where all vectors are described in *BFCS*. $m$ is the mass of the structure. The equations above have implied the center-of-gravity (COG) as the motion reference point. In the equations, $k = 1$ and $k = 2$ denote the 1st and 2nd order solutions, respectively. $\boldsymbol{F}_g$ and $\boldsymbol{M}_g$ are the force and moment vector, respectively. The subscript '$g$' indicates that the loads are calculated with respect to COG. $\boldsymbol{I}_B$ is the inertia matrix for rotations around COG, which is time-independent in a *BFCS* reference frame. In an *IneCS* reference frame, $\boldsymbol{I}_B$ may change with time (e.g., see Shao, 2010).

One should note that the second terms on the left-hand sides of Equation (22) are 2nd order, which are also kept when solving the 1st order hydrodynamic problem. We found those nonlinear body-motion equations provide more accurate results at the resonance of pitch/roll motions of ships.

Hydrodynamic loads $\boldsymbol{F}_g^{(k)}$ and $\boldsymbol{M}_g^{(k)}$ on a floating structure also include inertia parts associated with the added mass, thus care must be taken in the time-domain solution of the body-motion equations if an explicit time-integration scheme is applied. In this study, we have followed a stable approach proposed by Kim et al. (2008). In this approach, the forces and moments proportional to infinite-frequency added mass terms are added on both sides of the motion equations, thus the right-hand sides of the resulting Equations. (22a) and (22b) do not explicitly depend on the accelerations $\dot{\boldsymbol{v}}^{(k)}$ and $\dot{\boldsymbol{\omega}}^{(k)}$. Similar approaches have been applied in Kring (1994) and Huang (1997).

## 6. LOW-PASS FILTER BASED ON OPTIMIZED WEIGHTED LEAST SQUARES (WLS)

The landmark paper by Savitzky and Golay (1964) presented a smoothing algorithm based on least-squares to suppress random noises contained in the measured signal. The smoothing filters they described (hereafter referred to as SG filters) have been extensively applied in various engineering fields because they are versatile and easy to implement. When the data points are equally spaced, analytical solutions to the least-square equations can be found, in the form of convolution coefficients that can be applied to all sub-sets (also called stencils later in this paper) of the data, to obtain an estimation of the smoothed signal and its derivatives at the central point of each sub-set. The SG filters have been successfully applied, among others, by Engsig-Karup et al. (2009), Ducrozet et al. (2014), and Tong et al. (2021), in modeling fully nonlinear water waves. Despite it is becoming more and more popular in time-domain simulations related to ocean waves as well as wave-structure interactions, it is worth mentioning that the SG filters cannot eliminate the saw-tooth instabilities. Inspired by the original work of Savitzky and Golay (1964), novel low-pass filters with improved smoothing effects will be developed in this paper, based on the optimized weighted least-squares (WLS).

## 6.1 WLS-BASED FILTER

Suppose that we have a stencil consisting of $M_p$ points $\boldsymbol{x}_j = (x_j, y_j)$ with $j = 1, \cdots, M_p$ on the free surface, and $f_j = f(\boldsymbol{x}_j)$ is a datum value on $\boldsymbol{x}_j$. The data may contain undesirable saw-tooth noises that one wants to eliminate. An approximation of the exact function $f(\boldsymbol{x})$ may be assumed by the following series expansion

$$\hat{f}(\boldsymbol{x}) = \sum_{i+j \leq N_m} c_{i,j} x^i y^j. \tag{23}$$

Here both $i$ and $j$ are integers. $i + j$ is considered as the order of the polynomial $x^i y^j$. $N_m$ is a prescribed maximum value for $i + j$. The total number of terms in Equation (23) is $M = \frac{1}{2}(N_m + 1)(N_m + 2)$. Thus, Equation (23) can be reshaped into a 1D form as

$$\hat{f}(\boldsymbol{x}) = \sum_{s=1}^{M} a_s p_s(\boldsymbol{x}). \tag{24}$$

We will consider the following objective function

$$\mathcal{H}(a_1, \cdots, a_M) = \sum_{j=1}^{M_p} w_j [\hat{f}(\boldsymbol{x}_j) - f_j]^2, \tag{25}$$

where $w_j, j = 1, \cdots, N_p$, is the weighting coefficient for the $j$th data point. If $M_p > M$, an over-determinant system can be formed, leading to a least-squares problem. The least-squares solution to minimize $\mathcal{H}(a_1, \cdots, a_M)$ requires

$$\frac{\partial \mathcal{H}}{\partial a_s} = 0, \quad s = 1, \cdots, M, \tag{26}$$

which, after some algebra, gives the solution to $a_k$ as

$$\boldsymbol{a} = (a_1, \cdots, a_M) = \{\mathbf{B}^T \mathbf{W} \mathbf{B}\}^{-1} \mathbf{B}^T \mathbf{W}. \tag{27}$$

Here $\mathbf{B}$ is a matrix with entries of $b_{i,j} = p_j(\boldsymbol{x}_i)$ for $i = 1, \cdots, N_p$ and $j = 1, \cdots, M$. $\mathbf{W} = diag\{w_1, \cdots, w_{N_p}\}$ is a diagonal matrix. The vector $f = (f_1, \cdots, f_{N_p})^T$ contains the data at all discrete points.

Therefore, $f(\boldsymbol{x})$ can be approximated by Equation (24) as

$$\hat{f}(\boldsymbol{x}) = \boldsymbol{a}^T \boldsymbol{p} = y^T [\{\mathbf{B}^T \mathbf{W} \mathbf{B}\}^{-1} \mathbf{B}^T \mathbf{W}]^T \boldsymbol{p}, \tag{28}$$

where $\boldsymbol{p} = (p_1(x), \cdots, p_M(x))^T$. If we further define

$$q = [\{\mathbf{B}^T \mathbf{W} \mathbf{B}\}^{-1} \mathbf{B}^T \mathbf{W}]^T \boldsymbol{p}, \tag{29}$$

the approximated $f(\boldsymbol{x})$ in Equation (28) can be calculated as a linear superposition of the data set $f = (f_1, \cdots, f_{N_p})^T$ using the weighting coefficients defined in Equation (29), i.e.

$$\hat{f}(\boldsymbol{x}) = f^T \boldsymbol{q}. \tag{30}$$

The above derivation is based on data fitting using weighted least squares. However, Equation (30) can also be used as a low-pass filter. In particular, if an equidistant grid and unit weights $w_j = 1, j = 1, \cdots, N_p$, are concerned, the analytical solution to the convolution coefficients $\boldsymbol{c}$ can be obtained. Some examples of those have been tabulated in the original paper of Savitzky and Golay (1964), and some errors in the tables were corrected later in Steinier et al. (1972). The commonly used SG filters often take the following convolution form

$$\hat{f}_j = \sum_{i=j-\frac{1}{2}(M_p-1)}^{j+\frac{1}{2}(M_p-1)} c_i f_{j+i}. \tag{31}$$

The corresponding transfer function of the filter in Equation (28) and (31) for waves with wave number $\boldsymbol{k}$ is

$$G(\boldsymbol{k}) = \sum_{i=j-\frac{1}{2}(M_p-1)}^{j+\frac{1}{2}(M_p-1)} c_j e^{i \boldsymbol{k} \cdot \boldsymbol{x}_j}. \tag{32}$$

## 6.2 OPTIMIZATION OF WLS FILTER

As will be seen shortly, the SG filters do not necessarily eliminate the numerical saw-tooth waves, which have wavelengths of $2\Delta s$. Here $\Delta s$ is the local grid size on the free surface. However, it is possible to optimize the weighting coefficients of the WLS filters described in Section 6.1, and thus improve the performance of the low-pass filters.

Naturally, the points closer to the stencil center will be more important in the local approximation, thus their weighting coefficients $w_j$ should be larger than those further away from the stencil center. To achieve that, we define the following Gaussian function

$$g(\boldsymbol{x}) = \sqrt{\frac{6}{\pi D^2}} \exp\left\{-6 \frac{|\boldsymbol{x}|^2}{D_0^2}\right\}, \tag{33}$$

and the weighting coefficients based on the Gaussian function

$$w_j(\boldsymbol{x}_j; \boldsymbol{x}_0) = \frac{g_j(x_j - x_0)}{\sum_{j=1}^{N_p} g_j(x_j - x_0)}, \quad j = 1, \cdots, M_p. \tag{34}$$

Here $\boldsymbol{x}_0$ is the center of the stencil where the filtering or smoothing will be applied. $D_0$ is twice the local element size $\Delta h$, which is known for a given stencil. It can be understood as the local saw-tooth wavelength on a grid point. In this paper, $\Delta h$ is estimated as the minimum distance from the stencil points (excluding stencil center) to the stencil center, and thus $D_0$ is defined as

$$D_0 = \min_{j=1,\cdots,N_p; \, x_j \neq x_0} \{2|\boldsymbol{x}_j - \boldsymbol{x}_0|\}. \tag{35}$$

The parameter $D$ in Equation (33) is used to tune the shape and the influence area of the Gaussian function. When $D/D_0$

approaches infinity, all weighting coefficients will tend to be the same, i.e. $w_j \to 1/M_p$ as $D/D_0 \to \infty$ for $j = 1, \cdots, M_p$. $D/D_0$ can be optimized so that the modulus of $G(x)$ is minimized for saw-tooth waves, i.e. at $k_0 = 2\pi/D_0$.

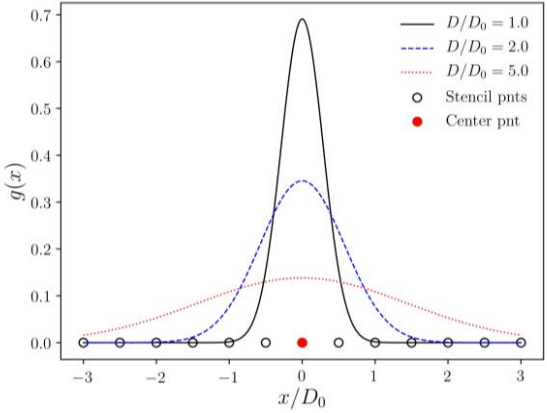

**FIGURE 5.** Gaussian weighting function for $D/D_0 = 1, 2, 5$ and an example of stencil consisting of 13 points with the center point in the middle.

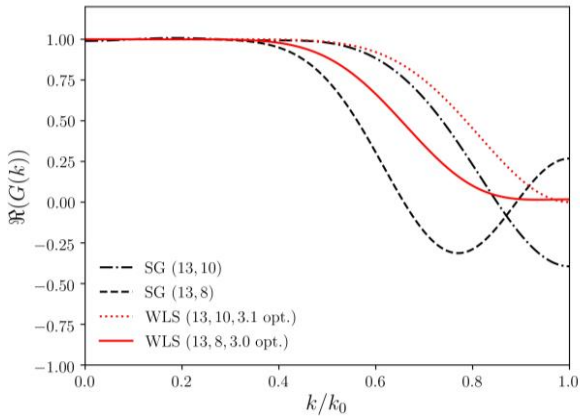

**FIGURE 6.** The real part of the transfer function $\Re(G(k))$ for 1D least-square-based filters and the corresponding optimized WLS filters. SG $(13, n)$ represents least squares with 13 stencil nodes and $n$th order polynomials. WLS $(13, n, D/D_0 \text{ opt.})$ means optimized WLS filter.

Figure 5 presents the Gaussian weighting function $g(x)$ for $D/D_0 = 1, 2, 5$, together with an example of a 13-point central stencil. The real part of the transfer function $\Re(G(k))$ for the 1D 13-point least-squares (LS) filters and the optimized WLS filters are compared in Figure 6. SG $(13, n)$ represents the SG filters with 13 nodes and $n$th order. WLS$(13, n, D/D_0 \text{ opt.})$ means optimized WLS filter, where $D/D_0$ opt. is the optimized ratio between $D$ and $D_0$ leading to minimum $|G(k)|$ for the saw-tooth waves. It is seen from the figure that the $\Re(G(k))$ curves for SG filters show oscillatory behavior for shorter waves, and they do not eliminate the saw-tooth waves denoted by $k/k_0 = 1$. SG filters may also lead to negative $\Re(G(k))$, suggesting that the filters may be too strong for some waves. On the contrary, the optimized WLS filters show decreasing $\Re(G(k))$ for shorter waves, and almost eliminate all energy of the saw-tooth waves. In general, the present optimal WLS filters are much less diffusive for longer waves that one wants to remain in the numerical solution.

To compare the efficiency of the SG and optimized WLS filters, a manufactured signal is constructed for $x \in [0,2]$ based on the following expression

$$f(x) = \sin(2\pi x) + 0.5\sin(\pi x/\Delta x), \quad (36)$$

Here $\Delta x$ is the size of discretization along $x$-axis. The 2nd term on the right-hand side is a noise with a period of $2\Delta x$. Figure 7 shows the raw signal, filtered results based on SG $(13, 10)$ filter, the optimized WLS $(13, 10)$ filter, and the analytical signal without the noise. Note that neither SG nor WLS filter can work perfectly at two ends of the signal, thus we have applied the selective filter by Berland et al. (2007) at the two ends. It is immediately clear that the new optimized WLS filter performs much better than the original SG filter, and is capable of eliminating the noise in the middle part of the signal.

In Figure 8, we present a 2D stencil as part of unstructured meshes. The 2D optimized WLS filter will be applied on the stencil center marked by the filled circle. The real part of the transfer function of the filter, i.e. $\Re(G(k))$, is presented in Figure 9 for both SG filters (denoted as LS filter in the figure) and the optimized WLS filters. In this figure, $(43, 5)$ stands for 43 stencil points and up to 5th order polynomials (see the definition in Equation (23)) are used in the construction of the filters. Again, better performances are seen for the optimized WLS filters.

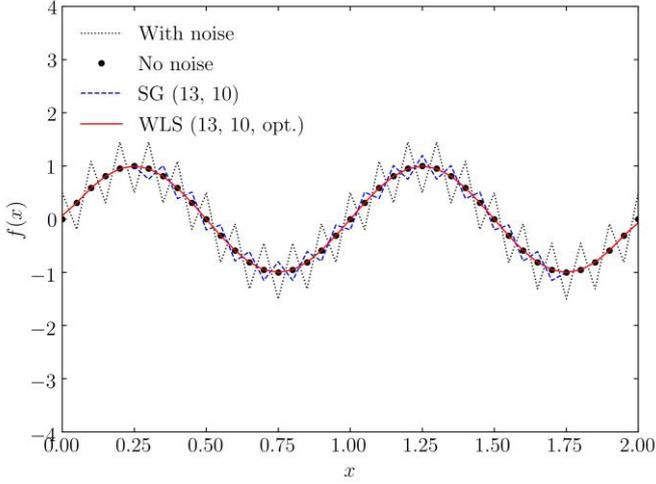

**FIGURE 7**. Comparison of the performance of the SG (13,10) filter and optimized WLS (13,10) filter for manufactured data with saw-tooth noises.

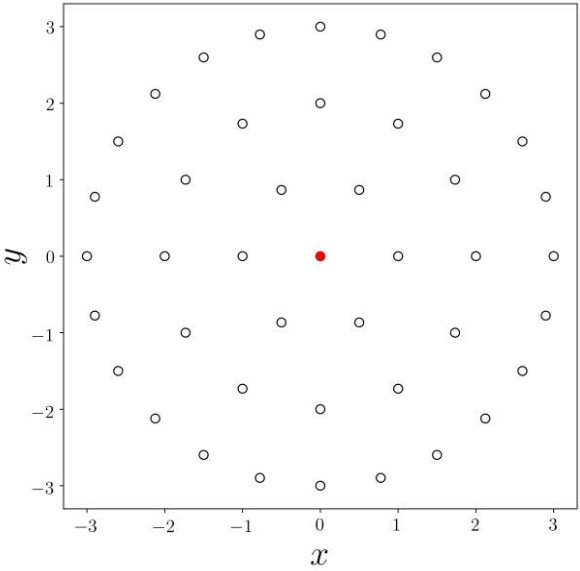

**FIGURE 8**. Example of stencil points and stencil center on the $xy$-plane. Filled circle: stencil center. Open circle: stencil points.

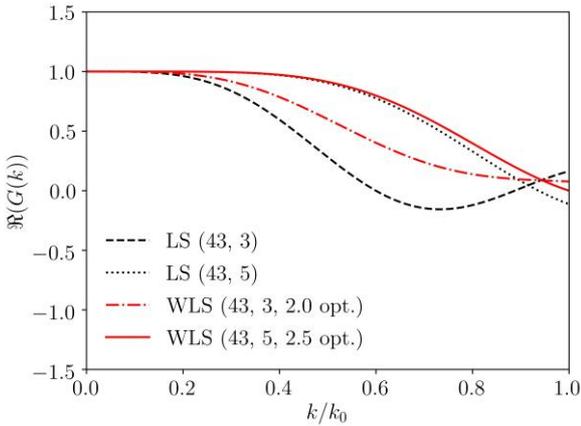

**FIGURE 9**. The real part of the transfer function $\Re(G(k))$ for 2D least-square-based filters and the corresponding optimized WLS filters. LS ($n1$, $n2$) represents least squares with $n_1$ stencil nodes and $n_2$ order polynomials. WLS ($13, n, D/D_0$ opt.) means optimized WLS filter.

In all the time-domain analyses in the next sections, similar optimized 2D filters to the WLS(43, 5) filter will be applied for the velocity potential and wave elevation on the free surface, except for the waterline, where the 1D filters are applied. To construct the stencil at each free-surface grid point, three surrounding layers of grid points are included. There are two options to apply the filters. One may directly apply the filters with an interval of, for instance, 5 to 30 time steps. Alternatively, the following mild filter, revised from Equation (31), can be applied at each time step

$$\hat{f}_j = (1-\alpha)f_j + \alpha \sum_{i=j-\frac{1}{2}(M_p-1)} c_i f_{j+i}. \quad (37)$$

Here $\alpha \in [0, 1]$ is a constant, typically taken as a much smaller value than 1.0. The second option is preferred in our analyses in this paper, and in our later numerical examples in Section 7, $\alpha = 0.05 \sim 0.1$ for the first 2 or 3 layers of points close to the waterline and $\alpha = 0.1 \sim 0.3$ for points away from the waterline, will be applied.

## 7. PRACTICAL APPLICATIONS

### 7.1 LINEAR SHIP SEAKEEPING AND ADDED RESISTANCE

One of the major tasks in ship hydrodynamic analysis is to calculate the WF motion responses and added resistance. Since there is no hydrostatic restoring in the surge, sway, and yaw, linearly increasing (with time) horizontal motions are expected in time-domain solutions for floating structures without station-keeping systems. This is associated with the homogeneous solution of the body-motion equations. Taking Newton's 2nd law for a single-degree-of-freedom system as an example, which can be formally expressed as $m\ddot{x} = f(t)$. Its homogeneous solution $x(t) = a_0 + a_1 t$ is in general not zero, where the constants $a_0$ and $a_1$ depend on the initial conditions. Thus, the corresponding velocity of the homogeneous solution is a constant $\dot{x} = a_1$. An example of such drifting in the surge is depicted in Figure 10 for the KVLCC2 ship traveling at $F_r = U/\sqrt{gL} = 0.142$ in a head-sea regular wave. Here $U$ is the forward speed of the ship, and $L$ is the ship length. The corresponding surge velocity is also shown in Figure 10.

If a linear or weakly-nonlinear formulation is developed in an *IneCS* reference frame, the numerical solution may eventually diverge due to ever-increasing horizontal motions. This is due to the so-called $m$-terms in the body-boundary condition, which include a contribution of $\left[x^{(k)} \cdot \nabla (\nabla \phi^{(0)})\right] \cdot \boldsymbol{n}$ due to the use of Taylor expansion when formulating the body boundary conditions. $x^{(k)}$ is the unsteady motion on the body surface, including contribution from both horizontal and vertical motions. Since the body boundary conditions will be applied in the solution of the Laplace equation, an excessively large horizontal motion will eventually destroy the time-domain numerical solutions. On the other hand, the occurrence of large unsteady motions violates the assumption behind a Taylor expansion. Therefore, artificial soft-springs are commonly applied in the numerical model to prevent the continuous drifting of the structure in the horizontal plane (Kring, 1994; Seo et al., 2014; Kim et al., 2017). Small artificial damping may also be applied for those horizontal motions. The stiffness of soft-spring and the strength of the damping should be defined with care, such that the LF transient horizontal motions have negligible effects on the WF responses.

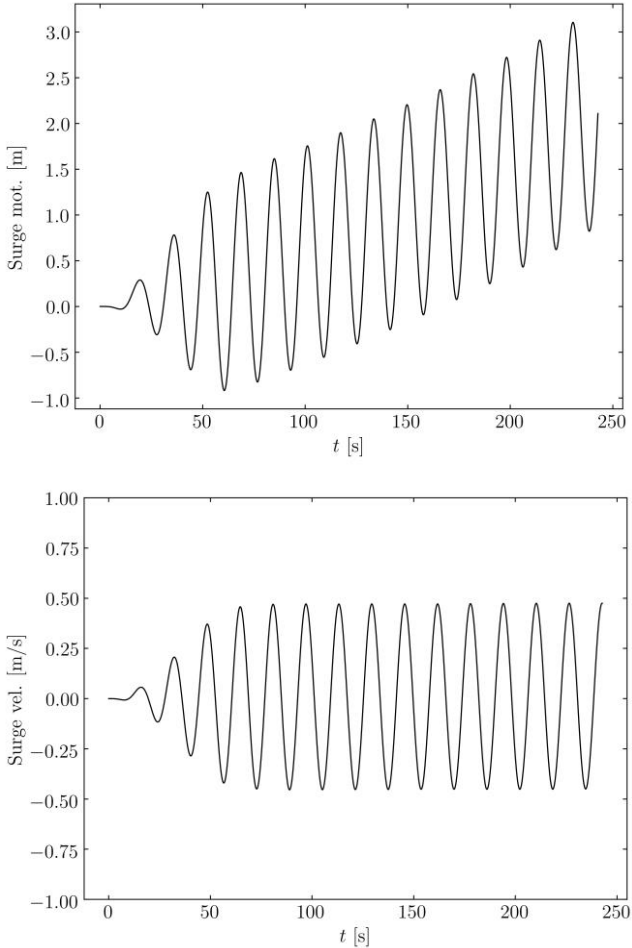

**FIGURE 10**. Time series for surge displacement and velocity of the KVLCC2 ship under a regular incident wave with $A = 3$ m and $\lambda/L = 2.0$. Top: surge motion, Bottom: surge velocity.

In the present numerical model described in previous sections, it is not necessary to use any artificial soft-springs to prevent the drifting in surge and sway. Due to the use of *BFCS* as the reference frame, the model does not involve similar terms as the $m$-terms in body boundary conditions. The free-surface conditions also do not involve the unsteady surge and sway displacements, but rather the velocities. In practice, it is much easier to minimize the velocities contributed by the homogeneous solution of the motion equations than the corresponding displacements, as the latter is an accumulation of the former in time. In this study, we use a simple but effective strategy, namely to apply a ramp function over a long duration when imposing the body boundary condition in the time domain, so that the transient effects are minimized. Numerical experiments for the KVLCC2 ship show that a ramp duration of 3 to 5 periods is enough for both long and short waves. The surge motion and velocity shown in Figure 10 are based on a ramp duration of 4 wave periods. As seen in the figure, the mean value of the surge velocity, i.e. the contribution of the homogeneous solution, is negligibly small compared with the dynamic part of the velocity. Since surge motion does not explicitly enter the solution of the boundary-value problem, its increase in time does not cause divergence of the numerical solution. The WF part of the surge motion can be easily extracted from the time series in a post-process through a standard high-pass filter.

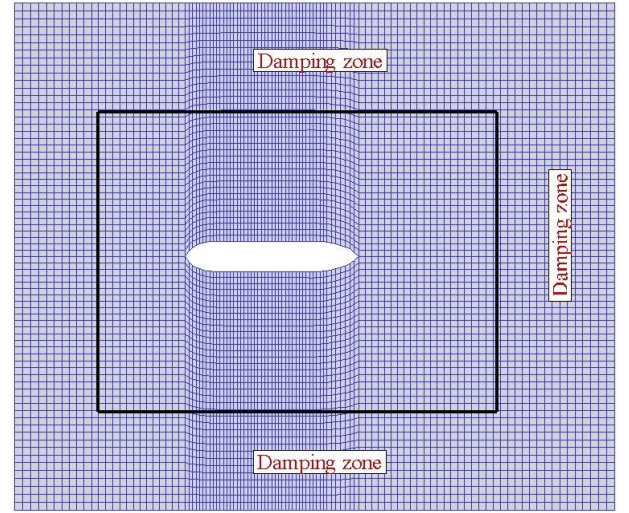

**FIGURE 11**. An illustration of the free surface meshes around the KVLCC2 ship. A damping zone at the outer layer of the free surface is also shown.

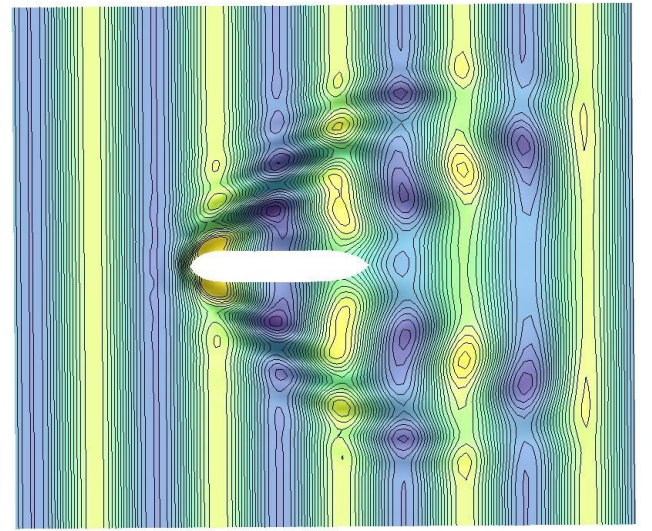

**FIGURE 12**. An example of a contour plot of the total unsteady velocity potential on the free surface for the KVLCC2 ship in head-sea regular waves with $\lambda/L = 0.7$ and $F_r = U/\sqrt{gL} = 0.142$.

**TABLE 2.** Main parameters for the KVLCC2 ship in full scale.

| Length, $L$ | 320 m |
|---|---|
| Breadth, $B$ | 58 m |
| Draft, $D$ | 20.8 m |
| Displ., $V$ | 312,622 m³ |
| LCG , fwd+ | 3.48% |
| VCG (%) | 16.58 m |
| Gyration radius ($r_{xx}, r_{yy}$) | $0.4B$, $0.25L$ |

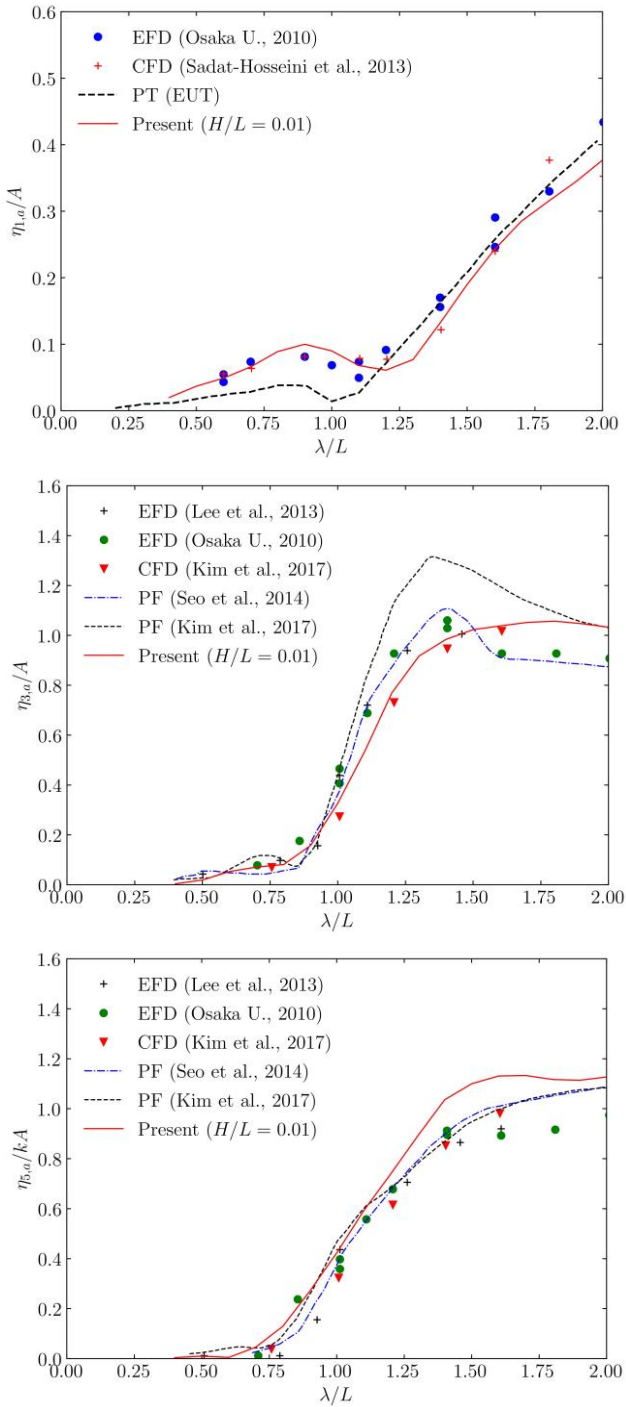

**FIGURE 13**. Surge, heave, and pitch motion RAOs for the KVLCC2 ship in head-sea regular waves. $F_r = 0.142$ is considered. Top: surge, Middle: heave, Bottom: pitch.

In the rest of this section, seakeeping and added resistance will be studied for the KVLCC2 ship, for which extensive validation and verification materials are available in the literature. The main particulars in full scale for the ship are listed in Table 2. A bird-view of the free-surface mesh model is shown in Figure 11. The computational domain is truncated on the free surface at one ship length ($L$) ahead of the bow and $2L$ on the two transverse sides as well as behind the stern. In total, 864 cubic elements (4548 nodes) on the free surface and 160 cubic elements (864 nodes) on the wetted ship surface are used in the discretization. A damping zone has been applied at the outer layer of the free surface. An example of the contour plot of the total (incident + scatter) unsteady velocity potential on the free surface is also presented in Figure 12 for $\lambda/L = 0.7$. For the plotting purpose, each 12-node cubic element (see e.g. Figure 3) has been split into nine 4-node elements in the post process.

The response amplitude operators (RAOs) for surge, heave, and pitch of the KVLCC2 in head-sea regular waves are presented in the top, middle, and bottom panels of Figure 13 respectively, while Figure 14 presents the corresponding added resistance of the ship. $F_r = 0.142$ has been considered in the analysis. The present results are compared with several model tests, 3D potential flow (PF), and 3D Computational Fluid Dynamic (CFD) results, with generally good agreement. Those reference results are manually extracted from Sadat-Hosseini et al. (2013), Seo et al. (2014), and Kim et al. (2017).

The 3D CFD models solving Navier-Stokes equations are still very time-consuming and need significant computational resources on clusters or supercomputers. Compared with other time-domain potential-flow models, the present time-domain model seems to be easier from a numerical modeling point of view, as the stiffness of the soft-spring must be designed so that its influence on other important dynamic responses is negligible.

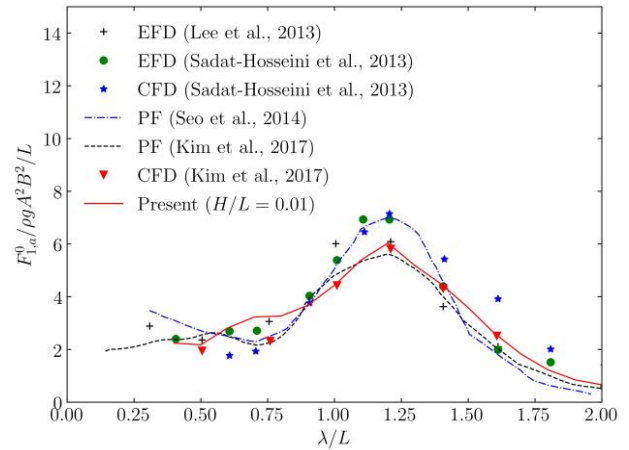

**FIGURE 14**. The non-dimensional added resistance of the KVLCC2 ship in head-sea regular waves. $F_r = 0.142$.

## 7.2 A FLOATING MONOPILE IN BI-CHROMATIC WAVES

In this subsection, the hydrodynamic responses of a moored floating monopile are studied in bi-chromatic waves. Due to the 2nd order nonlinear effects, the floating monopile will experience LF wave excitation loads, which are in general small in magnitudes. However, since the wave-radiation damping is negligibly small at resonant periods (typically in the order of 60 seconds or higher) of the horizontal motions, the 2nd order LF responses can be excessively large.

The main particulars of the considered floating monopile are taken from a spar Floating Offshore Wind Turbine (FOWT), as summarized in full scale in Table 3. The natural period for surge motion is approximately 76 seconds, and 26 seconds for heave, and 43 seconds for pitch. Some more details of the spar FOWT can also be found in Yang et al. (2021) where model tests in both regular and irregular waves have been carried out, and Zheng et al. (2020) where the main focus has been on the current effects on the 2nd order mean-drift and sum-frequency loads.

**TABLE 3.** Main particulars of the floating monopile taken from a spar FOWT design.

| Draft | 76 m |
| --- | --- |
| Mass | 11,137 t |
| Displ., | 11,420 m$^3$ |
| VCG | -49.0 m |
| Gyration radius ($r_{xx}, r_{yy}$) | 61.4 m |

The considered floating monopile is free to surge, heave and pitch under bi-chromatic incident waves, whose linear wave periods are $T_1 = 15$ s and $T_2 = 12.5$ s, respectively. The corresponding wave lengths are $\lambda_1 \approx 350$ m and $\lambda_2 \approx 244$ m, respectively. The considered wave steepness is $kA = 0.2$ for both wave components. The difference-frequency of the two wave components is $\Delta\omega = |\omega_1 - \omega_2| = 0.0838$ rad/s, almost the same as the natural frequency of the surge motion. $\omega_1$ and $\omega_2$ are the linear wave frequencies of the two waves, respectively.

The hydrodynamic mesh model of the considered floating monopile is shown in Figure 15. As illustrated at the top subplot, the free surface domain is truncated at a radius of 1.2 kilometers, corresponding to about 3.4 times the longer linear wave length $\lambda_1$, and a damping zone of more than $2\lambda_1$ at the outer layer of the domain is applied to absorb outgoing waves. A damping zone size from one to three wave lengths is quite typical in a similar type of analysis. The structure surface below the mean water surface is shown in the lower part of the figure. Note that different scales have been used in the two plots for a better illustration. To properly model the local waves close to the structure, we have used finer meshes near the waterline both on the body surface and the free surface. In Figure 15, around 30 cubic elements per wavelength close to the waterline are used. The final mesh model used in this analysis consists of 1932 cubic elements and 9703 nodes. A constant time step of $\Delta t = T_1/400$ is used to achieve convergent results for both linear and 2nd order results.

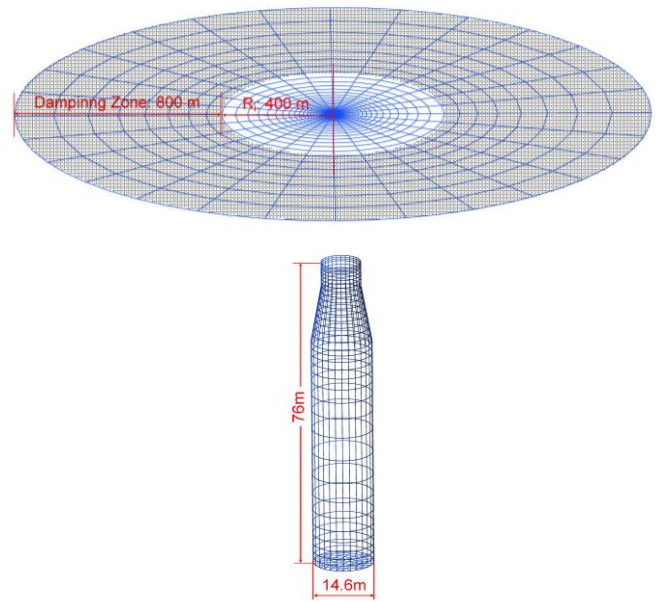

**FIGURE 15**. Mesh models of the floating monopile, whose geometry is taken from the floating foundation of a 6 MW spar-type FOWT in Yang et al. (2021). Top: free surface; Bottom: body surface.

The effects of large-amplitude horizontal motions and the 2nd order LF velocities are taken into account in the present time-domain analysis. At each time step, a more accurate description of the incident wave field around the structure is obtained at the instantaneous horizontal position, without using Taylor-expansion approximation around the static-equilibrium position of the structure, which is different from the traditional models. The 2nd order LF velocities of the structure are extracted by a low-pass filter based on Discrete Fourier Transform (DFT). This is needed as the complete 2nd order results include not only the difference-frequency ($|\omega_1 - \omega_2|$) quantities, but also the sum-frequency ($2\omega_1$, $2\omega_2$, $\omega_1 + \omega_2$) components. The LF velocities will be considered as quasi-steady speeds and used as inputs in the next time step of the time-domain simulation. More specifically, they are used to update $W^{(0)}$ and the associated $W^{(k)}$, $k = 1, 2$, in Equations (5) and (6). Therefore, its effect is also similarly taken into account as we have done for the forward speed effects of the KVLCC2 ship in Section 7.1. One difference is that the LF velocities of the floating monopile are solved as part of the time-domain solution, while the forward speed of the ship was prescribed. If the sum-frequency velocities are also used as inputs in Equations (5) and (6) in the next time step, the time-domain solution is less stable. On the other hand, since the local steady flow due to forward speed effects are modeled in our model by the double-body flow approximation, it is also questionable to apply a highly oscillatory forward speed in this model. As a final remark, since the sum-frequency responses are 2nd order quantities, the ignored effects are of 3rd order if we do not take sum-frequency velocities as a part of the quasi-steady velocity $W^{(0)}$. This can easily be understood by examining the order of the 2nd terms in Equations (5) and (6).

To partly illustrate the importance of the large horizontal motions, two different options of describing the incident wave field will be introduced. Their features are summarized in Table 4 and will be explained here. When describing the incident waves in either *BFCS* or *IneCS* reference frame, the effects of both the large-amplitude horizontal displacement and the corresponding velocity should be properly accounted for. As the first option, denoted as 'Present' in Table 4, the 1st and 2nd order incident waves fields around the structure are evaluated at the exact horizontal position due to both the WF and LF motions, while the influences of the horizontal LF velocities, e.g. on the encounter frequency of the waves, are also accounted for. In the second option, denoted as 'Traditional' in Table 4, the incident waves around the structure are approximated by Taylor expansion with respect to the mean positions, thus the 1st and 2nd order velocity potentials are obtained by $\phi_i^{(1)} = \phi_i^{(1)}|_{mean}$ and $\phi_i^{(2)} = \phi_i^{(2)}|_{mean} + x_1^{(1)}\phi_{i,x} + x_2^{(1)}\phi_{i,y}$, respectively. Here $\phi_i^{(k)}, i=1,2$ is the $k$th order velocity potential of incident waves. A subscript 'mean' indicates that the variable is evaluated at the mean horizontal position. $\phi_{i,x}$ and $\phi_{i,y}$ are the derivatives of $\phi_i^{(1)}$ with respect to $x$ and $y$, respectively. The second option is a commonly used approach in the traditional formulations based on *IneCS*.

**TABLE 4.** Summary of two different options to describe the 1st and 2nd order incident wave fields. The first option evaluates the incident waves at the instantaneous horizontal position based on WF + LF motions, while the second option approximates the incident waves through Taylor expansion around mean positions.

|  | Horizontal disp. | Horizontal vel. |
| --- | --- | --- |
| Present | Exact, WF+LF | LF |
| Traditional | Taylor exp., WF | None |

The time series of the 1st and 2nd order components of the motions (surge, heave, and pitch) for the floating monopile in the specified bi-chromatic waves are presented in Figures 16 and 17, respectively. The results based on the present and the traditional descriptions of the incident waves are presented. The simulations have been run for 1300 seconds, but only a duration of 300 seconds is presented, where steady states have been reached. In general, for the considered incident waves and the floating monopile in Table 4, large differences are seen for the two ways of describing the incident waves. For surge and pitch motions, the 1st order responses are more influenced than the corresponding 2nd order responses. However, the 2nd order heave response seems to have been more influenced by the large horizontal motions. The total (1st order + 2nd order) surge, heave, and pitch motions are also presented in Figure 18.

Here we have discussed the present and traditional ways of describing the incident waves and their consequences when applied in our numerical model described in the previous sections. One should not mistakenly understand that only the Froude-Krylov loads have been influenced in the analyses. Different incident wave descriptions lead to different solutions of the diffracted waves, rigid-body motions, and 2nd order wave-wave interaction between the incident, diffracted, and radiated waves. On the other hand, both the `Present' and `Traditional' results shown in this section were based on *BFCS* formulation, and the diffraction and radiation problems are solved at the instantaneous position of the horizontal position in both cases. Thus, our `Traditional' results in Figures 16, 17, and 18 cannot represent the results based on traditional formulation in *IneCS*. Nevertheless, the present results clearly demonstrate the importance of consistently accounting for the large horizontal motions for the studied floating monopile.

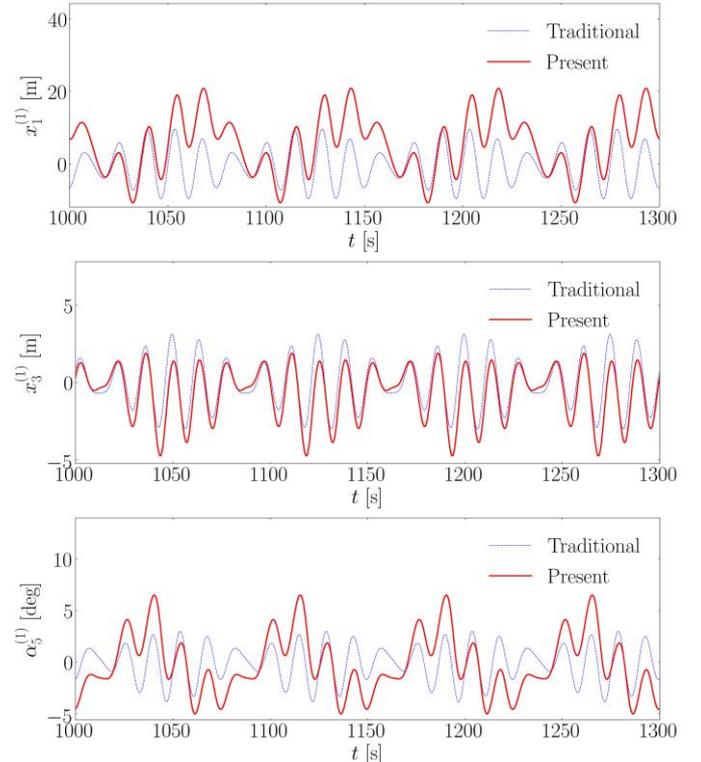

**FIGURE 16.** Time series of 1st order motion component of the floating monopile in bi-chromatic waves. Top: surge; Middle: heave; Bottom: pitch.

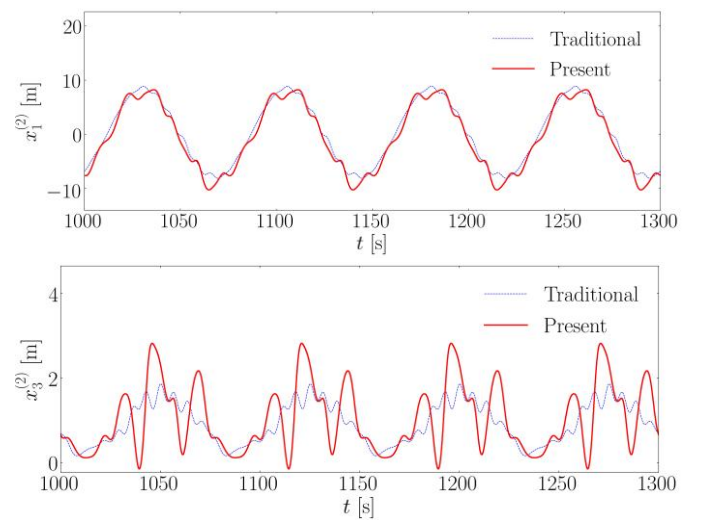

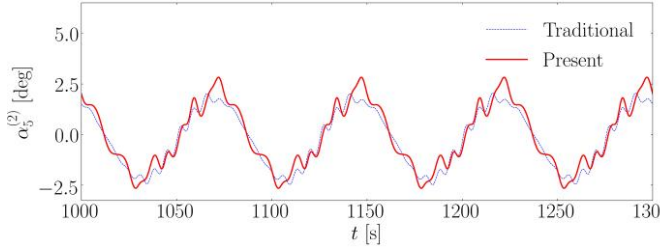

**FIGURE 17**. Time series of 2nd order motion component of the floating monopile in bi-chromatic waves. Top: surge; Middle: heave; Bottom: pitch.

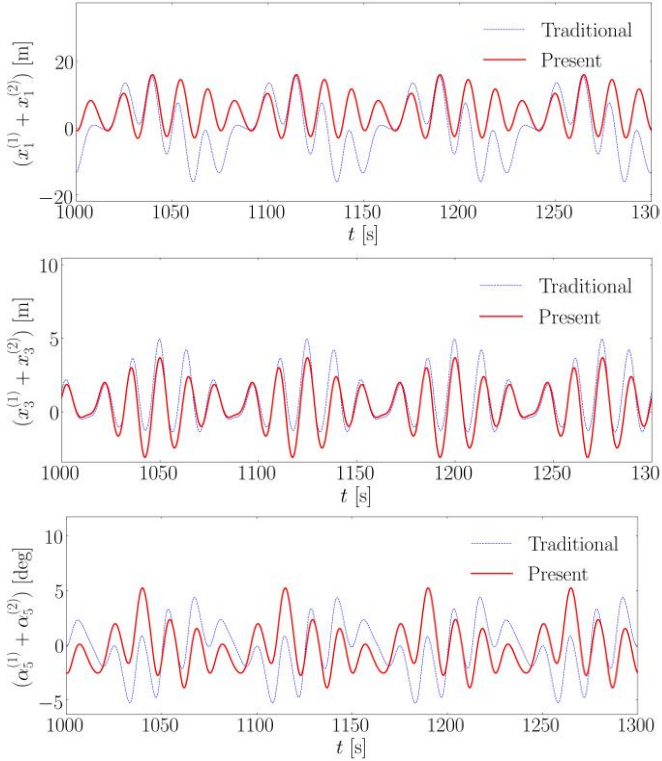

**FIGURE 18**. Time series of the total (1st order + 2nd order) motion of the floating monopile in bi-chromatic waves. Top: surge; Middle: heave; Bottom: pitch.

## 8. CONCLUSION

We present a complete and consistent time-domain computational model to analyze the 1st and 2nd order wave loads and the motions of floating structures, with particular emphasis on consistently accounting for the large horizontal displacements and velocities due to, for instance, the slow-drift motions of floating structures. The mathematical model based on a body-fixed reference frame is extended for this purpose, where the 2nd order low-frequency velocities are extracted at each time step and utilized as inputs of quasi-steady speeds of the structure in the next time step.

Another important advantage of the model is that it does not involve higher derivatives in either linear or 2nd order body-boundary conditions, which are however present in conventional formulations. A recently developed new set of explicit time-integration schemes for convection equations approximated from the corresponding implicit schemes. Contrasting with many other explicit schemes, which must be applied together with the upwind finite difference for spatial discretization, these schemes allow for the use of a central-difference type of scheme. A matrix-based linear stability analysis has been presented as a theoretical proof of the stability of a 2nd order explicit scheme that was used in the numerical studies in this paper.

To eliminate the weak saw-tooth waves on the free surface, we have also developed a new class of low-pass filters based on optimized weighted-least-squares, which are applicable for both structured and unstructured meshes. Direct comparison with the well-known 1D Savitzky-Golay filters and 2D least-squares filters which are a direct extension of the original Savitzky-Golay filters in two dimensions, clearly demonstrate that the new filters are more effective in eliminating short waves and less diffusive for longer waves.

Two practical examples have been studied to demonstrate the advantages of the developed computational model: seakeeping and added resistance of the KVLCC2 ship, and motions of a floating monopile in bi-chromatic waves. In the first example, we show that our model does not have to use soft-springs in surge and sway, while they are needed in the current best practices for a similar time-domain analysis. The second example clearly illustrates the importance to include the large horizontal motions in both the 1st and 2nd order analyses. As a final note, the present hydrodynamic model or similar should be used together with stochastic approaches (e.g. Naess and Moan, 2012; Glavind et al., 2021) to accurately predict the wave loads and structural responses in random waves.

## APPENDIX.

The 1st order and 2nd order transformation matrices $\mathcal{R}^{(1)}_{i \to b}$ and $\mathcal{R}^{(2)}_{i \to b}$ are defined as

$$\mathcal{R}^{(1)}_{i \to b} = \begin{bmatrix} 0 & \alpha^{(1)}_6 & -\alpha^{(1)}_5 \\ -\alpha^{(1)}_6 & 0 & \alpha^{(1)}_4 \\ \alpha^{(1)}_5 & -\alpha^{(1)}_4 & 0 \end{bmatrix}, \tag{38}$$

$$\mathcal{R}^{(2)}_{i \to b} = \begin{bmatrix} 0 & \alpha^{(2)}_6 & -\alpha^{(2)}_5 \\ -\alpha^{(2)}_6 & 0 & \alpha^{(2)}_4 \\ \alpha^{(2)}_5 & -\alpha^{(2)}_4 & 0 \end{bmatrix}$$
$$-\frac{1}{2} \begin{bmatrix} \left(\alpha^{(1)}_5\right)^2 + \left(\alpha^{(1)}_6\right)^2 & 0 & 0 \\ -2\alpha^{(1)}_4 \alpha^{(1)}_5 & \left(\alpha^{(1)}_4\right)^2 + \left(\alpha^{(1)}_6\right)^2 & 0 \\ -2\alpha^{(1)}_4 \alpha^{(1)}_6 & -2\alpha^{(1)}_5 \alpha^{(1)}_6 & \left(\alpha^{(1)}_4\right)^2 + \left(\alpha^{(1)}_5\right)^2 \end{bmatrix}. \tag{39}$$